\documentclass
[nofootinbib,preprint,12pt,prd,groupedaddress,showpacs,eqsecnum,noshowkeys,titlepage]{revtex4}%
\usepackage{amsfonts}
\usepackage{amsmath}
\usepackage{amssymb}
\usepackage{graphicx}
\usepackage{latexsym}%
\begin{document}
\title{Radion clouds around evaporating black holes }
\author{J.R. Morris}
\affiliation{Physics Dept., Indiana University Northwest, 3400 Broadway, Gary, Indiana
46408, USA}
\email{jmorris@iun.edu}

\begin{abstract}
A Kaluza-Klein model, with a matter source associated with Hawking radiation
from an evaporating black hole, is used to obtain a simple form for the radion
effective potential. The environmental effect generally causes a
matter-induced shift of the radion vacuum, resulting in the formation of a
radion cloud around the hole. There is an albedo due to the radion cloud, with
an energy dependent reflection coefficient that depends upon the size of the
extra dimensions and the temperature of the hole.

\end{abstract}

\pacs{11.27.+d, 4.50.Cd, 98.80.Cq}
\keywords{nontopological soliton, radion cloud, black hole albedo}\maketitle

\section{Introduction}

Kaluza-Klein type models involving compactified extra dimensions produce
effective 4d theories containing moduli fields (radions) that are associated
with scale factors of the compact dimensions. Particle masses typically
exhibit a radion dependence, and local matter sources contribute to an
effective potential for these scalar fields. If the compactification is
\textit{inhomogeneous}, particle masses and charges can have spatial and
temporal variations. Soliton-like structures associated with these scalars may
result (see, for example, \cite{DG1,DG2,db1,db2,db3}), with possibly
observable consequences. An extreme condition is considered here, where the
matter density from an evaporating microscopic black hole (MBH) may become
large enough to give rise to an inhomogeneous compactification, resulting in a
radionic cloud around the black hole. The reflective properties of this cloud
endow the near horizon region with a radion induced albedo.

Scenarios of this type involving a radion coupled to the radiative field
around an evaporating black hole have been studied previously in refs.
\cite{GSS} and \cite{JM}. In \cite{GSS} rather general conditions were
considered, and in \cite{JM} attention was focused on finite temperature,
one-loop quantum corrections due to a partially thermalized medium surrounding
an evaporating black hole. The resulting effective potential is quite
complicated, and difficult to represent in a simple closed form. Here,
however, I take a more classical approach and consider a simple, but explicit,
class of potentials $V(\varphi)$ for the radion field $\varphi$. This type of
radion potential was studied first by Davidson and Guendelman in
\cite{DG1,DG2} in the context of a Freund-Rubin compactification, and later by
Carroll, Geddes, Hoffman, and Wald (CGHW) in \cite{CGHW}, using an extra
dimensional two-form magnetic field. Both treatments result in the same radion
potential for the case of two extra dimensions, but the former treatment also
holds for an arbitrary number $n$ of extra dimensions. In either case, the
resulting potential has a rather simple form for an assumed set of parameter
relations. For simplicity and concreteness, I specialize to the case of two
extra dimensions, and refer to the radion potential $V(\varphi)$ as the
DG-CGHW radion potential, although its generalization for arbitrary $n,$ given
by \cite{DG1,DG2}, is also given below. In addition to functional simplicity,
the potential allows the extra dimensions to be stabilized classically,
without explicit quantum corrections.

Using this simple DG-CGHW model, the full effective radion potential
$U(\varphi)$ can be developed which includes a matter-sourced correction due
to the Hawking radiation matter field. The matter source contribution to the
radion effective potential $U(\varphi)$ depends, in a simple way, upon the
local matter density $\rho(\mathbf{r},t)$ and radion mass $m_{\varphi}$. A
knowledge of these parameters then allows, in principle, a determination of
the spatial and temporal variation of the radion field in the vicinity of the
evaporating hole. The matter contribution can induce a shift, or complete
destabilization, of the radion vacuum near the horizon, resulting in a
\textquotedblleft radion cloud\textquotedblright\ around the MBH. This cloud
has an associated energy-dependent reflectivity, which can result in a
distortion of the infrared portion of the Hawking radiation spectrum, as well
as a partial reflection of low energy particles incident upon the MBH from the
outside. By looking at the forms of the effective potential $U$ near the
hole's horizon and at asymptotic distances, it is suggested that the radion
cloud has an evolving size $R$ $\lesssim m_{\varphi}^{-1}$, and a maximal
reflection coefficient $\mathcal{R}_{\max}$ that depends upon the matter
density $\rho_{hor}$ near the horizon through the ratio $\rho_{hor}%
/(m_{\varphi}^{2}M_{0}^{2})$, where $M_{0}=1/\sqrt{8\pi G}$ is the reduced
Planck mass.

\section{Radion effective potential}

\subsection{Effective 4d action}

We start by considering a $D=(4+n)$ dimensional spacetime having $n$ compact
extra spatial dimensions endowed with a metric given by
\begin{equation}
ds_{D}^{2}=\tilde{g}_{MN}dx^{M}dx^{N}=\tilde{g}_{\mu\nu}(x)dx^{\mu}dx^{\nu
}+b^{2}(x^{\mu})\gamma_{mn}(y)dy^{m}dy^{n} \label{e1}%
\end{equation}

where $x^{M}=(x^{\mu},y^{m})$. Here $M,N=0,1,2,3,\cdot\cdot\cdot,D-1$ label
all the spacetime coordinates, while $\mu,\nu=0,1,2,3$, label the 4d
coordinates, and $m,n$ label those of the compact extra space dimensions. The
extra dimensional scale factor $b(x^{\mu})$ is assumed to be independent of
the $y$ coordinates and takes the form of a scalar field in the 4d effective
theory. The extra dimensional metric $\gamma_{mn}(y)$ depends upon the
geometry of the extra dimensional space and is related to $\tilde{g}%
_{mn}(x,y)$ by $\tilde{g}_{mn}=b^{2}\gamma_{mn}$. As in
refs.\cite{DG1,DG2,CGHW}, consideration is restricted to extra dimensional
compact spaces with constant curvature, with a curvature parameter $k$ defined
by%
\begin{equation}
k=\frac{\tilde{R}[\gamma_{mn}]}{n(n-1)} \label{e1a}%
\end{equation}

The action for the $D$ dimensional theory is%
\begin{equation}
S_{D}=\int d^{D}x\sqrt{\left\vert \tilde{g}_{D}\right\vert }\left\{  \frac
{1}{2\kappa_{D}^{2}}\left[  \tilde{R}_{D}[\tilde{g}_{MN}]-2\Lambda\right]
+\mathcal{\tilde{L}}_{D}\right\}  \label{e2}%
\end{equation}

where $\tilde{g}_{D}=\det\tilde{g}_{MN}$, $\tilde{R}_{D}$ is the Ricci scalar
built from $\tilde{g}_{MN}$, $\Lambda$ is a cosmological constant for the $D$
dimensional spacetime, $\mathcal{\tilde{L}}_{D}$ is a Lagrangian for the
fields in the $D$ dimensions, $\kappa_{D}^{2}=8\pi G_{D}=V_{y}\kappa^{2}%
=V_{y}8\pi G$, where $G(G_{D})$ is the 4d ($D$)-dimensional gravitational
constant, and $V_{y}=\int d^{n}y\sqrt{\left\vert \gamma\right\vert }$ is the
coordinate \textquotedblleft volume\textquotedblright\ of the extra
dimensional space. A mostly negative metric signature $(+,-,-,\cdot\cdot
\cdot,-)$ is used here.

The action can be expressed in terms of an effective 4d action (see, for
example,\cite{CGHW,NDL-JM} for details) which takes the form%
\begin{equation}%
\begin{array}
[c]{ll}%
S & =%
{\displaystyle\int}
d^{4}x\sqrt{-\tilde{g}}\left\{  \dfrac{1}{2\kappa^{2}}[b^{n}\tilde{R}%
[\tilde{g}_{\mu\nu}]-2nb^{n-1}\tilde{g}^{\mu\nu}\tilde{\nabla}_{\mu}%
\tilde{\nabla}_{\nu}b-n(n-1)b^{n-2}\tilde{g}^{\mu\nu}(\tilde{\nabla}_{\mu
}b)(\tilde{\nabla}_{\nu}b)\right. \\
& \left.  +n(n-1)kb^{n-2}]+b^{n}\left[  \mathcal{L}_{D}-\dfrac{\Lambda}%
{\kappa^{2}}\right]  \right\}
\end{array}
\label{e7}%
\end{equation}

in the 4d Jordan frame (with metric $\tilde{g}_{\mu\nu}$), and I have defined
a normalized field Lagrangian, $\mathcal{L}_{D}=V_{y}\mathcal{\tilde{L}}_{D}$.
A 4d Einstein frame metric $g_{\mu\nu}$ can be defined:%
\begin{equation}
\tilde{g}_{\mu\nu}=b^{-n}g_{\mu\nu},\ \ \ \ \ \tilde{g}^{\mu\nu}=b^{n}%
g^{\mu\nu},\ \ \ \ \ \sqrt{-\tilde{g}}=b^{-2n}\sqrt{-g} \label{e8}%
\end{equation}

The action $S$ in (\ref{e7}), in terms of the 4d Einstein metric, takes the
form%
\begin{equation}%
\begin{array}
[c]{ll}%
S & =%
{\displaystyle\int}
d^{4}x\sqrt{-g}\left\{  \dfrac{1}{2\kappa^{2}}\left[  R[g_{\mu\nu}%
]+\dfrac{n(n+2)}{2}b^{-2}g^{\mu\nu}(\nabla_{\mu}b)(\nabla_{\nu}%
b)+n(n-1)kb^{-(n+2)}\right]  \right. \\
& \ \ \ \ \ \ \ \ \ \ \ \ \ \ \ \ \ \ \ \left.  +b^{-n}\left[  \mathcal{L}%
_{D}-\dfrac{\Lambda}{\kappa^{2}}\right]  \right\}
\end{array}
\label{e9}%
\end{equation}

where total derivative terms have been dropped. Furthermore, an effective 4d
source, or matter, Lagrangian $\mathcal{L}_{m}$ can be defined in terms of the
$D$ dimensional source Lagrangian $\mathcal{L}_{D}=V_{y}\mathcal{\tilde{L}%
}_{D}$ and the scale factor $b$:%
\begin{equation}
\mathcal{L}_{m}=b^{-n}\mathcal{L}_{D} \label{e10}%
\end{equation}

A scalar radion field $\varphi$ with a canonical kinetic term is now defined
by%
\begin{equation}
\sqrt{\frac{n(n+2)}{2\kappa^{2}}}\ln\frac{b}{b_{0}}=\varphi,\ \ \ \ b=b_{0}%
\exp\left(  \sqrt{\frac{2}{n(n+2)}}\ \kappa\varphi\right)  \label{e11}%
\end{equation}

where $b_{0}$ is some constant, which will be set equal to unity, and
$\kappa=\sqrt{8\pi G}=\sqrt{8\pi}/M_{P}=M_{0}^{-1}$ is the inverse of the
reduced Planck mass.

\subsection{Radion potential $V$}

For a simple and concrete example, I use the radion potential studied in
refs.\cite{DG1,DG2,CGHW} for the case of $n=2$ extra dimensions having a
constant positive curvature parameter $k$. I also adopt the same choices of
parameter relations to obtain a simple functional form. The potential $V$ has
contributions from the curvature term in (\ref{e9}), the cosmological constant
term $\Lambda$, and either (1) an extra dimensional magnetic field due to
$F_{45}=\sqrt{|\gamma|}F_{0}$, where $F_{0}$ is a constant, in $\mathcal{L}%
_{D}$ (see\cite{CGHW}), or (2) a Freund-Rubin term (see \cite{DG1,DG2}) of the
form%
\begin{equation}
\mathcal{\tilde{L}}_{D}=-\frac{1}{48}F^{2},\ \ \ F_{MNPQ}=\partial_{\lbrack
M}A_{NPQ]},\ \ \ F_{\mu\nu\lambda\sigma}=\sqrt{\lambda}\frac{\sqrt{|\tilde
{g}|}}{3b^{n}}\varepsilon_{\mu\nu\lambda\sigma} \label{e11a}%
\end{equation}
The potential $V$ obtained by CGHW for the case of $n=2$ extra dimensions and
an extra dimensional magnetic field $F_{45}$ leads to a simple potential
(written here in terms of the scale factor $b$, rather than in terms of the
radion field $\varphi$) given by%
\begin{equation}
V(b)=\lambda\left(  b^{-6}-2b^{-4}+b^{-2}\right)  \ \ \ \ \ \text{(CGHW
potential)} \label{e12}%
\end{equation}

with the following relations and parameter choices: (see eqs. (36) and (37) in
ref.\cite{CGHW})%
\begin{equation}
\varphi=\frac{2}{\kappa}\ln b,\ \ \ b=e^{\frac{1}{2}\kappa\varphi
},\ \ \ \ \lambda=\frac{k}{2\kappa^{2}}=\frac{m_{\varphi}^{2}M_{P}^{2}}{16\pi
}=\frac{1}{2}m_{\varphi}^{2}M_{0}^{2} \label{e13}%
\end{equation}

where $k=\frac{\partial^{2}V(\varphi)}{\partial\varphi^{2}}\Big|_{\varphi
=0}=m_{\varphi}^{2}$ is the mass$^{2}$ of the radion field\cite{CGHW}, $M_{P}$
is the Planck mass, and $M_{0}=1/\kappa=M_{P}/\sqrt{8\pi}$ is the reduced
Planck mass. The form of $V$ (see Figure 1; $V$ is also sketched in
refs.\cite{DG1} and \cite{CGHW}) has one local minimum, which for the case
$n=2$ is located at $b=1$ (or $\varphi=0$), with $V(b=1)=0$, followed by a
barrier for $b>1$, then an asymptotic decrease with $V\rightarrow0$ as
$b\rightarrow\infty$.%

\begin{center}
\includegraphics[
trim=0.000000in 0.000000in 0.002166in -0.003799in,
height=4.8965cm,
width=8.2979cm
]%
{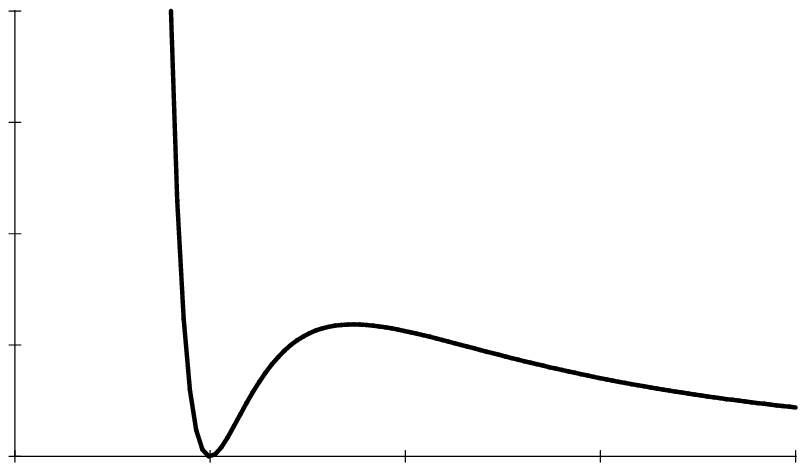}%
\\
{}%
\end{center}

{\small Figure 1: A plot is shown of }$V(b)/\lambda$ {\small  vs }$b${\small .
The local minimum occurs at  }$b=1$ {\small  (}$\varphi=0${\small ) where
}$V=0${\small . The local maximum occurs at  }$b\approx1.75$ {\small  where
}$V/\lambda\approx.15${\small . As }$b\rightarrow\infty${\small ,
}$V\rightarrow0${\small .}

\medskip

For the more general case of $n$ extra dimensions given by \cite{DG1,DG2}, the
Davidson-Guendelman (DG) potential arising from the Freund-Rubin term above is
given by (with factors of $\kappa$ reinstated here)%
\begin{equation}
V(b)=\dfrac{\lambda}{b^{3n}}-\frac{1}{2}\frac{k}{\kappa^{2}}n(n-1)\dfrac
{1}{b^{n+2}}+\dfrac{\Lambda/\kappa^{2}}{b^{n}}\text{\ \ \ \ \ (DG potential)}
\label{e13a}%
\end{equation}

For the case of $n=2$, requiring the potential to vanish at its minimum leads
to the condition%
\begin{equation}
\Lambda=\frac{k^{2}/\kappa^{2}}{4\lambda} \label{e13b}%
\end{equation}

Imposing the parameter choice $\lambda=k/2\kappa^{2}$ as in (\ref{e13}) then
gives the same potential as in (\ref{e12}) with a local minimum at $b=1$ where
$V=0$. For simplicity and concreteness, attention is restricted here to this
simple form of the potential $V$ \ for the particular case of $n=2$ extra
dimensions. For our case of $n=2$, the potential $V$, given by (\ref{e12}) and
(\ref{e13}), has a local minimum at $b=1$ ($\varphi=0$) where $V=0$, a local
maximum located at $b>1$, and the potential falls off exponentially,
$V\rightarrow0$ as $b\rightarrow\infty$.

An equation of motion (EoM) is obtained from (\ref{e9}) for the radion field
$\varphi$,%
\begin{equation}
\square\varphi+\frac{\partial V}{\partial\varphi}-\left\langle \frac
{\partial\mathcal{L}_{m}}{\partial\varphi}\right\rangle =0 \label{e14}%
\end{equation}

where $\mathcal{L}_{m}$ is the matter Lagrangian which depends upon scalar,
spinor, and vector matter fields, as well as the radion field. Using field
redefinitions, the $\varphi$ dependence of $\mathcal{L}_{m}$ appears in
particle masses $m_{A}(\varphi)$ and in gauge coupling constants (see, e.g.,
\cite{DP}). The matter Lagrangian will therefore contribute to an effective
potential $U(\varphi)$ for the radion.

\subsection{Matter contribution--Hawking radiation}

The matter contribution to the EoM for $\varphi$ (or $b$) comes from the
Hawking radiation\cite{Hawking} from an evaporating black hole with surface
temperature $T$ (as seen asymptotically). Denote this matter contribution to
(\ref{e14}) by $\sigma=\left\langle \frac{\partial\mathcal{L}_{m}}%
{\partial\varphi}\right\rangle $. The Lagrangian $\mathcal{L}_{m}$ contains
the matter and gauge fields, such as fermionic terms like\cite{NDL-JM,DP}
$\mathcal{L}_{\psi}=\bar{\psi}(i\gamma\cdot\partial-m(\varphi))\psi$ with
$m(\varphi)=b^{-\frac{n}{2}}(\varphi)m_{0}$, ($m_{0}$ = const) and gauge field
terms, such as that for the photon, $\mathcal{L}_{F}=-\frac{1}{4}b^{n}%
(\varphi)F_{\mu\nu}F^{\mu\nu}$. For a freely propagating electromagnetic field
with $F_{\mu\nu}F^{\mu\nu}=0$, we have no contribution to $\sigma$ from
$\mathcal{L}_{F}$. However, particle modes with nonzero rest mass do
contribute to $\sigma$ through terms like $-\alpha\left\langle m\bar{\psi}%
\psi\right\rangle $ where%
\begin{equation}
\alpha_{A}=\frac{\partial\ln m_{A}(\varphi)}{\partial\varphi} \label{e15}%
\end{equation}

with $A$ labeling the particle species. (In the nonrelativistic flat space
limit this $\sigma$ term is proportional to the fermionic energy density,
$-\alpha\left\langle m\bar{\psi}\psi\right\rangle \sim-\alpha\rho$. However,
we want to consider $\sigma$ terms beyond the nonrelativistic flat space
limit. We will conclude that $\sigma\sim-\alpha g_{\mu\nu}\mathcal{T}%
_{cl}^{\mu\nu}=-\alpha\mathcal{T}_{cl}$ for the more general case where
$\mathcal{T}_{cl}^{\mu\nu}$ is a classical stress-energy tensor, and that,
classically, $\alpha_{A}=\alpha=$ const., independent of particle species.)

Now, for simplicity, rather than using the field theoretic version of the
matter Lagrangian $\mathcal{L}_{m}$, let us follow the approach used by Damour
and Polyakov\cite{DP} and treat the matter with a classical description,
replacing the field theoretic action with a classical particle action $S_{cl}%
$. We consider particle modes having a nonzero rest mass $m_{A}(\varphi)$ with
$\partial_{\varphi}\mathcal{L}_{A}\neq0$ and write a classical action%
\begin{align}
S_{cl}  &  =-\sum_{A}\int ds_{A}m_{A}=-\sum_{A}\int m_{A}\left[  g_{\mu\nu
}(x_{A})dx_{A}^{\mu}dx_{A}^{\nu}\right]  ^{1/2}\nonumber\\
&  =-\sum_{A}\int d^{4}x\int m_{A}\left[  g_{\mu\nu}(x_{A})dx_{A}^{\mu}%
dx_{A}^{\nu}\right]  ^{1/2}\delta^{(4)}(x-x_{A})=\int d^{4}x\sqrt
{-g}\mathcal{L}_{cl} \label{e16}%
\end{align}

and identify $\sqrt{-g}\mathcal{L}_{cl}=-\sum_{A}\int m_{A}\left[  g_{\mu\nu
}(x_{A})dx_{A}^{\mu}dx_{A}^{\nu}\right]  ^{1/2}\delta^{(4)}(x-x_{A})$. \ A
field theoretic energy-momentum tensor for the matter fields, defined by
$\mathcal{T}^{\mu\nu}=\dfrac{2}{\sqrt{-g}}\dfrac{\partial\left(  \sqrt
{-g}\mathcal{L}_{m}\right)  }{\partial g_{\mu\nu}}$, is replaced by an
energy-momentum tensor $\mathcal{T}_{cl}^{\mu\nu}$ for the classical
particles, with\footnote{With this set of definitions we have $\mathcal{T}%
_{00}>0$ for both fields and classical particles.}
\begin{equation}
\mathcal{T}_{cl}^{\mu\nu}=-\frac{2}{\sqrt{-g}}\frac{\partial\left(  \sqrt
{-g}\mathcal{L}_{cl}\right)  }{\partial g_{\mu\nu}}=\frac{1}{\sqrt{-g}}%
\sum_{A}\int m_{A}u_{A}^{\mu}u_{A}^{\nu}\delta^{(4)}(x-x_{A})d\tau_{A}
\label{e17}%
\end{equation}

where $u^{\mu}=dx^{\mu}/d\tau$ satisfies an \textquotedblleft
on-shell\textquotedblright\ constraint $g_{\mu\nu}u^{\mu}u^{\nu}=1$. Taking
the trace gives%
\begin{equation}
\mathcal{T}_{cl}=g_{\mu\nu}\mathcal{T}_{cl}^{\mu\nu}=\frac{1}{\sqrt{-g}}%
\sum_{A}\int m_{A}\delta^{(4)}(x-x_{A})d\tau_{A}=-\mathcal{L}_{cl} \label{e18}%
\end{equation}

We therefore find%
\begin{equation}
\sigma=\frac{\partial\mathcal{L}_{cl}}{\partial\varphi}=\sum_{A}\alpha
_{A}\mathcal{L}_{cl,A}=-\sum_{A}\alpha_{A}\mathcal{\mathcal{T}}_{cl,A}
\label{e19}%
\end{equation}

for $\alpha_{A}=$ const. The constant $\alpha_{A}$ takes a value $\alpha
_{A}=\alpha=-\sqrt{\frac{n}{2(n+2)}}\ \kappa$, which for our $n=2$ model
becomes $\alpha=-\kappa/2$. This can be seen\cite{Dicke} by considering the
matter action,%
\begin{equation}
S=-\sum_{A}\int m_{0},_{A}\ d\tilde{s}_{A} \label{new1}%
\end{equation}

where $m_{0},_{A}$ is the constant Jordan frame particle mass for species $A$,
and $d\tilde{s}=\sqrt{\tilde{g}_{\mu\nu}dx^{\mu}dx^{\nu}}$ is the Jordan frame
line element, which by (\ref{e8}), is related to the Einstein frame line
element $ds=\sqrt{g_{\mu\nu}dx^{\mu}dx^{\nu}}$ by $d\tilde{s}=b^{-\frac{n}{2}%
}ds$. The matter action rewritten in the Einstein frame is%
\begin{equation}
S=-\sum_{A}\int m_{0},_{A}\ (b^{-\frac{n}{2}}ds_{A})=-\sum_{A}\int
m_{A}\ ds_{A} \label{new2}%
\end{equation}

where the Einstein frame mass is%
\begin{equation}
m_{A}=b^{-\frac{n}{2}}m_{0},_{A}=\exp\left(  -\sqrt{\frac{n}{2(n+2)}}%
\ \kappa\varphi\right)  m_{0},_{A} \label{new3}%
\end{equation}

Eqs. (\ref{e15}) and (\ref{new3}) then give%
\begin{equation}
\alpha_{A}=\frac{\partial\ln m_{A}(\varphi)}{\partial\varphi}=-\sqrt{\frac
{n}{2(n+2)}}\ \kappa\rightarrow-\frac{\kappa}{2}\text{\ \ for \ \ }n=2
\label{new4}%
\end{equation}

From (\ref{e19}) we therefore have $\sigma=-\alpha\mathcal{T}_{cl}%
=\frac{\kappa}{2}\mathcal{T}_{cl}$ for our $n=2$ model.

The Hawking radiation (for a neutral nonrotating black hole) is assumed to be
a fluid with energy density $\rho=\mathcal{T}_{0}^{0}$ and a normal radial
pressure component $p_{r}=-\mathcal{T}_{r}^{r}$, and tangential
pressure\footnote{The spacetime is assumed to have spherical symmetry.}
$p_{T}=-\mathcal{T}_{\theta}^{\theta}=-\mathcal{T}_{\phi}^{\phi}$ in the
center of momentum frame, i.e., the rest frame of the black hole. We also
assume the energy density and pressures to be related by the equations of
state%
\begin{equation}
p_{r}=w_{r}\rho,\ \ \ p_{T}=w_{T}\rho,\ \ \ 0\leq w_{r}\leq1,\ \ \ 0\leq
w_{T}\leq1 \label{new4a}%
\end{equation}

where $w_{r,T}$ are constants. For an isotropic perfect fluid $p_{T}=p_{r}=p$
and $w_{r}=w_{T}=w$ with $0\leq w\leq1$. Let us define an effective pressure
$p$ and a parameter $w$ by
\begin{equation}
\ p=\frac{1}{3}(p_{r}+2p_{T}),\ \ \ \ w=\frac{1}{3}\left(  w_{r}%
+2w_{T}\right)  ,\ \ \ \ 0\leq w\leq1 \label{new4b}%
\end{equation}

so that an effective equation of state can be written in the form $p=w\rho$,
as assumed by Zurek and Page\cite{Zurek-Page} and by 't Hooft\cite{'t Hooft},
where the Hawking radiation is regarded as a perfect fluid with the constant
$w=p/\rho\in\lbrack0,1]$.

The trace of the stress-energy tensor becomes%
\begin{equation}
\mathcal{T}_{cl}=\rho-(p_{r}+2p_{T})=[1-(w_{r}+2w_{T})]\rho=(1-3w)\rho
\label{new4c}%
\end{equation}

For an ideal gas of noninteracting massless particles in thermal equilibrium,
$w=1/3$. However, we proceed by leaving $w$ as a free parameter, subject to
$0\leq w\leq1$ as assumed in \cite{Zurek-Page} and \cite{'t Hooft}. This
allows for a value of $\mathcal{T}_{cl}$ that can be positive, negative, or
zero. With (\ref{e19}) and (\ref{new4c}) we obtain our approximate result
\begin{equation}
\sigma=\frac{\partial\mathcal{L}_{cl}}{\partial\varphi}=-\alpha\mathcal{T}%
_{cl}=-\alpha(1-3w)\rho\label{e21}%
\end{equation}

with the energy density $\rho$ being dominated by relativistic particle modes.

A couple of remarks are in order here. First, we note that $\mathcal{T}_{cl}$
and therefore $\rho$ in (\ref{e21}) are generated by the particle modes with
nonzero rest mass and do not include the energy density $\rho_{0}$ and
pressure $p_{0}$ that is due to the massless (e.g. photon) components of the
radiation. The total energy density and pressure of the entire fluid would be
$\rho_{\text{tot}}=\rho+\rho_{0}$ and $p_{\text{tot}}=p+p_{0}$, respectively.
Secondly, it may be that there are multiple components of the fluid
corresponding to various particle species, with $\rho=\sum_{A}\rho_{A}$, and
each species may have an equation of state $p_{A}=p_{A}(\rho)$ which, in
principle, could be complicated. However, in order to study the effects of the
radiation, these difficulties are avoided here by our simple assumption that
$p/\rho=w=$ const., the same assumption made in the Hawking radiation fluid
models of refs.\cite{Zurek-Page} and\cite{'t Hooft}. This seems palatable for
a case where there are very few relativistic massive modes, at least on time
scales sufficiently small compared to the black hole evaporation time scale
$M/\dot{M}$.

The EoM \ $\square\varphi+\partial_{\varphi}V-\sigma=0$ \ for the radion field
$\varphi$ becomes%
\begin{equation}
\square\varphi+\frac{\partial V}{\partial\varphi}+\alpha(1-3w)\rho=0
\label{e22}%
\end{equation}

An effective potential $U(\varphi)$ is now defined by $U=V-\sigma
\varphi=V+U_{matter}$, or%
\begin{equation}
U(\varphi)=V(\varphi)+\alpha(1-3w)\rho\varphi=V(\varphi)+\frac{2}{\kappa
}\alpha(1-3w)\rho\ln b\text{\ \ \ \ (effective potential)} \label{e23}%
\end{equation}

where $\ln b=\frac{1}{2}\kappa\varphi$ for our model with two extra dimensions
and we set $U(\varphi=0)=0$. The matter contribution to the radion effective
potential is%
\begin{equation}
U_{matter}=-\sigma\varphi=-\frac{2}{\kappa}\sigma\ln b=\frac{2\alpha}{\kappa
}\mathcal{T}_{cl}\ln b=-\mathcal{T}_{cl}\ln b=-(1-3w)\rho\ln b \label{new6}%
\end{equation}
where the result $2\alpha/\kappa=-1$ from (\ref{new4}) has been used. The sign
of this matter term is controlled by the parameter $w$, and in the special
case $w=1/3$ then $\sigma\rightarrow0$ and matter term does not contribute to
the radion effective potential. We note that for $w<1/3$ then $U_{matter}$ is
a decreasing function for $b>1$, while for $w>1/3$ we have that $U_{matter}$
is an increasing function for $b>1$.

\subsection{Radion effective potential, $U$}

With (\ref{e12}), (\ref{e13}), (\ref{e23}), and (\ref{new6}) we can now write
an explicit, but simple, effective potential in the form
$U(b)=V(b)-\mathcal{T}_{cl}\ln b$, or
\begin{subequations}
\label{e29}%
\begin{align}
\frac{1}{\lambda}U(b)  &  =\left(  b^{-6}-2b^{-4}+b^{-2}\right)
-\frac{\mathcal{T}_{cl}}{\lambda}\ln b\label{e29a}\\
&  =\left(  b^{-6}-2b^{-4}+b^{-2}\right)  -\zeta\ln b \label{e29c}%
\end{align}

where%
\end{subequations}
\begin{equation}
\zeta\equiv\frac{\mathcal{T}_{cl}}{\lambda}=\frac{2\mathcal{T}_{cl}%
}{m_{\varphi}^{2}M_{0}^{2}}=\frac{(1-3w)\rho}{\lambda} \label{e30}%
\end{equation}
Here, the parameter $\zeta$ is dimensionless and $\zeta=\zeta(r,t)$ is a
function of radial distance $r$ from the black hole since $\rho=\rho(r,t)$.
The assumed range of $w$ allows a value of $\zeta$ in the range $-2\rho
/\lambda\leq\zeta\leq\rho/\lambda$.

Figure 2 gives a representation of $U(b)/\lambda$ for various
\textit{positive} values of $\zeta$ ($p\leq\rho/3$). The asymptotic vacuum
value of $b$ occurs at $b_{0}=1$ for $\zeta=0$, but for values of
$0<\zeta\lesssim.5$ the vacuum value of $b$ is shifted to larger values,
$b>1$, and the minimum of $U$ \ becomes more negative. For values
$\zeta\gtrsim.5$ the local minimum disappears, the vacuum state is completely
destabilized, and $U(\varphi)$ is a monotonically decreasing function whose
slope depends on $\zeta$.%

\begin{figure}
[h]
\begin{center}
\includegraphics[
height=4.8902cm,
width=8.3127cm
]%
{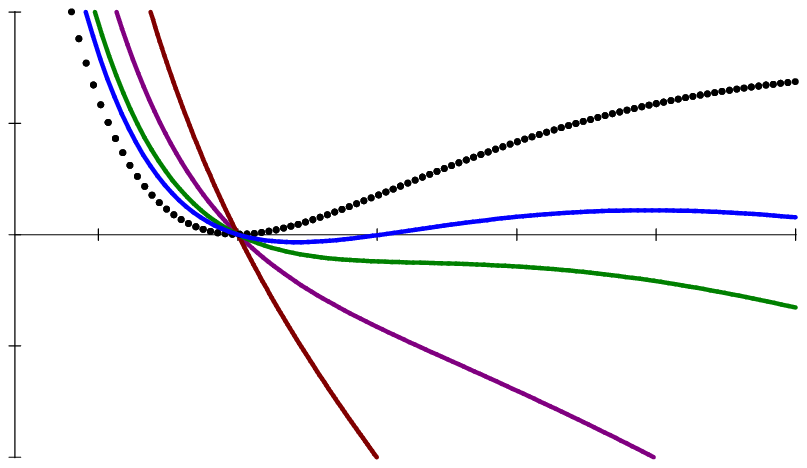}%
\end{center}
\end{figure}

{\small Figure 2: Plots of }$U(b)/\lambda$ {\small  vs. }$b$ {\small  are
shown for positive values of }$\zeta${\small . The dotted curve, with a
minimum at }$U=0$ {\small  and }$b=1${\small , has }$\zeta=0$ {\small  and
corresponds to the radion potential }$V(b)/\lambda${\small . The solid curves
have }$\zeta=.3,.5,1,2${\small , with the more negatively sloped curves
corresponding to bigger }$\zeta${\small . The vacuum value of }$b$ {\small
occurs at }$b_{0}=1$ {\small  for }$\zeta=0${\small , but for }$0<\zeta
\lesssim.5$ {\small  the vacuum value of }$b$ {\small  is shifted to larger
values, }$b_{vac}>1${\small , and the minimum of }$U$ {\small  completely
disappears for }$\zeta\gtrsim.5${\small .}

\bigskip

Figure 3 gives a representation of $U(b)/\lambda$ for various
\textit{negative} values of $\zeta$ \ ($p\geq\rho/3$). The asymptotic vacuum
value of $b$ occurs at $b_{0}=1$ for $\zeta=0$, but for larger values of
$|\zeta|$ the vacuum value of $b$ is shifted to smaller values, $b<1$, and the
minimum of $U$ becomes more negative.%

\begin{figure}
[h]
\begin{center}
\includegraphics[
trim=0.000000in 0.000000in -0.035115in -0.092605in,
height=4.9134cm,
width=8.2874cm
]%
{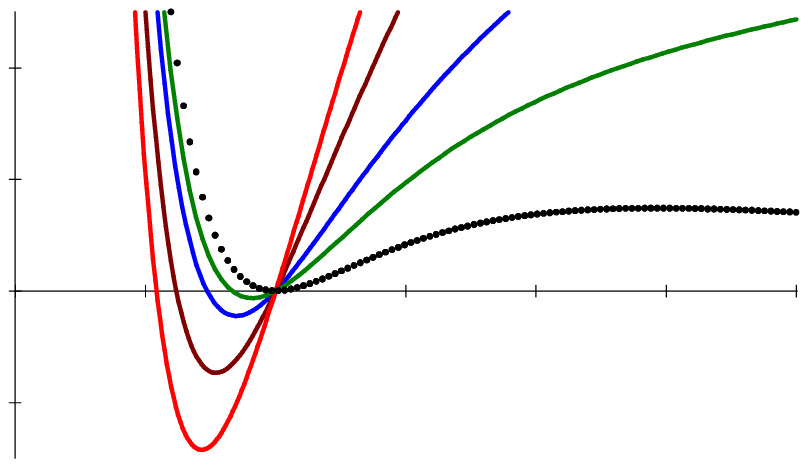}%
\end{center}
\end{figure}

{\small Figure 3:\ Plots of }$U(b)/\lambda$ {\small  vs. }$b$ {\small  are
shown for negative values of }$\zeta${\small . The dotted curve, with a
minimum at }$U=0$ {\small  and }$b=1${\small , has }$\zeta=0$ {\small  and
corresponds to the radion potential }$V(b)/\lambda${\small . The solid curves
have }$\zeta=-.5,-1,-2,-3${\small , with the lower minimum curves
corresponding to bigger }$|\zeta|${\small . The vacuum value of }$b$ {\small
occurs at }$b_{0}=1$ {\small  for }$\zeta=0${\small , but for larger values of
}$|\zeta|${\small  the vacuum value of }$b$ {\small  is shifted to smaller
values, }$b_{vac}<1${\small , and the minimum of }$U$ {\small  becomes more
negative.}

\bigskip

Thus the vacuum values $b_{vac}$ and $\varphi_{vac}$ become $r$ dependent in
general. Far from the hole, $\zeta\rightarrow0$ and $\varphi_{vac}%
\rightarrow0$, $b_{vac}\rightarrow1$. Near the hole, where $\zeta\neq0$, then
$\varphi\neq0$ $\ $and $b\neq1$. Therefore $\varphi$ interpolates between a
positive or negative value $\varphi\neq0$ near the horizon to $\varphi=0$ at
asymptotic distances. As the hole evaporates and $|\zeta|$ increases, any
vacuum state near the horizon gets further shifted to smaller or larger
values, depending on the sign of $\zeta$. For $\zeta>0$ a stable vacuum
eventually disappears and the radion rolls to larger values until the hole's
explosive end.

\section{Radion cloud and black hole albedo}

\subsection{Radion cloud}

An energy momentum tensor $\mathcal{T}_{\mu\nu}$ has been defined for the
matter portion $\mathcal{L}_{m}$ of the effective 4d Lagrangian%

\begin{equation}
\mathcal{L}=\mathcal{L}_{\varphi}(\varphi)+\mathcal{L}_{m}(\varphi,\sigma
,\psi,\cdot\cdot\cdot)=\frac{1}{2}(\partial\varphi)^{2}-V(\varphi
)+\mathcal{L}_{m}(\varphi,\sigma,\psi,\cdot\cdot\cdot) \label{e31}%
\end{equation}

and an energy momentum tensor $S_{\mu\nu}$ can be written for the pure radion
part $\mathcal{L}_{\varphi}$:%
\begin{equation}
S_{\mu\nu}=\partial_{\mu}\varphi\partial_{\nu}\varphi-g_{\mu\nu}%
\mathcal{L}_{\varphi}=\partial_{\mu}\varphi\partial_{\nu}\varphi-g_{\mu\nu
}\left[  \tfrac{1}{2}\partial^{\alpha}\varphi\partial_{\alpha}\varphi
-V(\varphi)\right]  \label{e33}%
\end{equation}

The energy density part for the radion field,%
\begin{equation}
S_{00}=\frac{1}{2}(\partial_{0}\varphi)^{2}-\frac{1}{2}g_{00}g^{rr}%
(\partial_{r}\varphi)^{2}+g_{00}V(\varphi) \label{e34}%
\end{equation}

vanishes asymptotically, but becomes nonzero near the evaporating black hole
where $\varphi$ develops a nonzero vacuum value $\varphi_{vac}$ due to the
environmental effects of a nonzero $\zeta$ in the effective potential $U$.

For a nonradiating black hole (i.e., a matter-vacuum solution with
$\mathcal{L}_{m}=0$, $T_{\mu\nu}=0$, $\sigma=0$) we have the Einstein equation
$R_{\mu\nu}-\frac{1}{2}g_{\mu\nu}R=-\kappa^{2}S_{\mu\nu}$ along with the
radion EoM $\square\varphi+V^{\prime}(\varphi)-\sigma=0$. The minimal energy
radion solution ($S_{\mu\nu}=0$) is given by the trivial solution $\varphi=0$
($b=1$). Therefore the gravitational field alone of the black hole has no
effect on the radion and there is no radion cloud in this case. However, for a
radiating black hole with nonzero values of $\sigma$ and $\zeta$ outside the
horizon, $\varphi=0$ is not a solution of the radion EoM and the radion field
$\varphi(r,t)$ must interpolate between a value of $\varphi_{hor}\neq0$ near
the horizon and $\varphi=0$ asymptotically. Since $\varphi\neq0$ is not at the
minimum of $V(\varphi)$, then $V(\varphi)>0$, contributing a positive
contribution to $S_{00}$. There are nonnegative gradient terms contributing to
$S_{00}$ as well. So near the horizon, $S_{00}>0$, and asymptotically
$S_{00}\rightarrow0$.

The energy density of the scalar field $\varphi$ is concentrated near the MBH,
where the gradient terms are large, and the radion field forms a cloud around
it. The exact structure of this cloud requires a knowledge of the solution
$\varphi(r,t)$ to the EoM\ \ \ $\square\varphi+U^{\prime}(\varphi)=0$. A crude
estimate of the extent of this cloud of energy can be obtained by considering
a thin shell of Hawking radiation at a radius $r\gg r_{S}$ where $r_{S}=2GM$
is the radius of a Schwarzschild black hole. (Relativistic radiation with
speed $v\sim1$ is assumed, as a higher energy density is carried by
relativistic modes.) At this asymptotic distance the mass $\delta M$ in the
spherical shell is approximately constant as it propagates outward, so that
$\delta M\approx4\pi r^{2}\rho(r,t)\delta r\sim P(t)\delta t$ where
$P(t)=-\dot{M}(t)>0$ is the power output of the radiation from the evaporating
MBH and for the sake of simplicity I have neglected any time retardation
effects in $P(t)$. This gives a crude estimate for the matter density%
\begin{equation}
\rho(r,t)\sim\frac{P(t)}{4\pi r^{2}} \label{b1}%
\end{equation}

At a fixed instant $t$ the matter density drops off as $r^{-2}$, while at a
fixed distance $r$ the density increases with time as described by the power
output $P(t)$. An outer \textquotedblleft edge\textquotedblright\ of the
evolving radion cloud, i.e. the cloud radius $R(t)$, can be defined as a
radial distance where the density $\rho$ assumes a sufficiently small constant
value, i.e., $\eta\equiv\rho/\lambda\ll1$ is a small constant, so that
$\rho\approx0$ outside this radius. From (\ref{b1}) this cloud radius, where
$\eta$ is a constant, is given by%
\begin{equation}
R^{2}(t)\sim\frac{P(t)}{4\pi\rho}=\frac{P(t)}{4\pi\eta\lambda}=\frac
{P(t)}{2\pi\eta m_{\varphi}^{2}M_{0}^{2}} \label{b2}%
\end{equation}

The ordinary Steffan-Boltzmann law for a perfect blackbody (no gravitational
greybody effects assumed, for simplicity) with emitting surface area $4\pi
r_{S}^{2}$ gives $P(t)=4\pi r_{S}^{2}\sigma_{SB}T^{4}(g/2)$ where
$g=(N_{B}+\frac{7}{8}N_{F})$ is the effective number of degrees of freedom of
relativistic particles and $\sigma_{SB}=\pi^{2}/60$ is the Steffan-Boltzmann
constant. Using $r_{S}=2GM=\frac{M}{4\pi M_{0}^{2}}$ and $T=\frac{1}{8\pi
GM}=\frac{M_{0}^{2}}{M}$ (where $M_{0}=1/\kappa=1/\sqrt{8\pi G}$ is the
reduced Planck mass) leads to%
\begin{equation}
P(t)=\frac{(g/2)\pi}{240}\frac{M_{0}^{4}}{M^{2}(t)} \label{b3}%
\end{equation}

From this, the cloud radius is given by%
\begin{equation}
R(t)\sim\sqrt{\frac{g/2}{480\eta}}\left(  \frac{M_{0}}{M(t)}\right)
m_{\varphi}^{-1}\sim\left(  \frac{M_{0}}{M(t)}\right)  m_{\varphi}^{-1}
\label{b4}%
\end{equation}

where, for simplicity, I have set $480\eta/(g/2)\sim1$ and $M(t)$ is the black
hole mass. The radion cloud grows in size as the hole shrinks, with $\dot
{R}/R\sim-\dot{M}/M$. From (\ref{b4}) $R$ approaches an upper limit $R_{\max
}\sim m_{\varphi}^{-1}$ near the explosive end of the MBH as $M\rightarrow
M_{0}$. The requirement that $r_{S}/R\ll1$ implies that $m_{\varphi}\ll
\frac{4\pi M_{0}^{3}}{M^{2}}\lesssim M_{0}$. Provided that the rather natural
condition $m_{\varphi}\ll M_{0}$ is satisfied, then so is the requirement that
$R\gg r_{S}$.

The energy dependent reflection coefficient (see below) $\mathcal{R}(\omega)$
and transmission coefficient $\mathcal{T}(\omega)=1-\mathcal{R}(\omega)$ for a
particle of energy $\omega$ will depend upon the variation in $\varphi$ (or
$b$) from the near-horizon region to the asymptotic region, along with the
radius $R$ of the cloud. The reflection coefficient approaches a maximal value
$\mathcal{R}_{\max}$ as particle energy $\omega$ approaches a minimal
value\cite{NDL-JM}.

\subsection{Radion reflectivity and black hole albedo}

Basic features expected of particle reflectivity by the radion field -- a
radion induced black hole albedo -- can be obtained from the (flat space)
results of \cite{NDL-JM} and other studies of particle reflection from
ordinary (nonradionic) scalar field domain walls (see, for
example,\cite{VSbook,Everett,EDW1,EDW2} ). In the case of photons, the radion
cloud, treated as a scalar modulus domain wall of thickness $\sim R$, has a
maximal reflection coefficient given by\cite{NDL-JM}%
\begin{equation}
\mathcal{R}_{\max}=\dfrac{\left(  b_{1}^{2}-b_{2}^{2}\right)  ^{2}}{\left(
b_{1}^{2}+b_{2}^{2}\right)  ^{2}}=\dfrac{\left(  b_{hor}^{2}-1\right)  ^{2}%
}{\left(  b_{hor}^{2}+1\right)  ^{2}} \label{a1}%
\end{equation}

where $b_{1}$, $b_{2}$ are the values of the scale factor $b$ on the two
different sides of a modulus wall. The two \textquotedblleft
sides\textquotedblright\ of our domain wall are the black hole horizon where
$b=b_{hor}$ and the asymptotic region approximately a distance $R$ away where
$b\rightarrow1$. The transmission coefficient is $\mathcal{T}(\omega
)=1-\mathcal{R}(\omega)$ and for photons of energy $\omega$ we have the thin
wall limit with $\mathcal{R}(\omega)\rightarrow\mathcal{R}(0)=\mathcal{R}%
_{\max}$ in the infrared limit $\omega\rightarrow0$, where the photon
wavelength $\lambda_{\gamma}\gg R$. From\cite{NDL-JM} it was found from
numerical calculations that $\mathcal{R}(\omega)/\mathcal{R}_{\max}$ typically
begins to become significant for energies $\omega\lesssim1/R$, i.e., the thin
wall limit. Note that $\mathcal{R}_{\max}\sim1$ when $b_{hor}\gg1$, or
$b_{hor}\ll1$. We may have $b_{hor}\gg1$ for $\zeta_{hor}\gtrsim.5$, at which
point the near-horizon vacuum of $U(b)$ completely destabilizes. For a
near-horizon value of $0<\zeta_{hor}<.5$, one expects $\mathcal{R}_{\max}\ll
1$, as there is a minimum of $U$ that is not far removed from $b=1$. On the
other hand, for negative $\zeta$ with $|\zeta|\gg1$ we have $b_{hor}%
\rightarrow0$ as $\zeta\rightarrow-\infty$, in which case $\mathcal{R}_{\max
}\rightarrow1$ again. Since $|\zeta|$ increases with black hole temperature
$T$, one expects $\mathcal{R}_{\max}$ to increase with increasing $T$. These
considerations lead us to expect a possible alteration of the infrared portion
of the transmitted Hawking radiation, as well as a partial reflection from
near-horizon regions of low energy particles incident upon the black hole from
outside. (The details of the spectral distortion, however, will depend upon
the structure of the radion cloud.) On the other hand, for high energy photons
with $\omega\gg1/R$, the cloud becomes transparent (thick-wall
limit)\cite{NDL-JM} with $\mathcal{R}\rightarrow0.$ Similar qualitative
statements are expected for massive particle modes. The black hole therefore
has an energy-dependent albedo associated with the radion cloud, which in
turn, is due to an inhomogeneous compactification of the extra dimensions near
the horizon.

The above deductions are based upon reflection and transmission
characteristics in flat space. The effects of curved space would alter the
gradient terms appearing in the $\square\varphi$ portion of the radion EoM,
and therefore the gradient nature of the solution $\varphi(r,t)$. The exact
expressions for $\mathcal{R}(\omega)$ and $\mathcal{T}(\omega)$ would depend
on the exact solution $\varphi$, but the basic qualitative features mentioned
above for $\mathcal{R}(\omega)$ are not expected to be significantly affected.

\subsection{$\mathbf{\zeta}$ \textbf{near the horizon}}

\ In the limit of a static, ideal fluid in thermodynamic equilibrium, the
local total energy density\footnote{Recall that $\rho_{\text{tot}}=\rho
+\rho_{0}$, or $\rho=f\rho_{\text{tot}},$ where $f=\rho/\rho_{\text{tot}%
}=1-\rho_{0}/\rho_{\text{tot}}$ is the fraction of energy carried by
nonmassless particles.} is\cite{Zurek-Page} $\rho_{\text{tot}}\sim
T^{\ast(w+1)/w}$, where $T^{\ast}(r)=T/\sqrt{g_{00}(r)}=(\sqrt{g_{00}}8\pi
GM)^{-1}$ is the blue-shifted Hawking temperature. In this limit
$\zeta=(1-3w)\rho/\lambda\sim(1-3w)T^{\ast(w+1)/w}/\lambda$ can become quite
large or divergent near the horizon (or would-be horizon). (There can be
significant backreactions on the metric, and the studies\cite{Zurek-Page}%
,\cite{'t Hooft} suggest that the horizon could be removed by a (static)
Hawking \textquotedblleft atmosphere\textquotedblright, with $\rho$ remaining
finite.) If the horizon is not removed by backreactions, the local energy
density can diverge on the horizon, due to the diverging blue-shifted local
temperature\cite{Barbon}. Furthermore, quantum field effects such as vacuum
polarization\cite{DFU},\cite{CF},\cite{FZ} are expected to play important
roles and may contribute to the $\sigma=\left\langle \partial\mathcal{L}%
_{m}/\partial\varphi\right\rangle $ term in the effective potential. In any
case, whether $\rho$ diverges or remains finite near the black hole, the local
value of $|\zeta|$ and $\sigma$ may become extremely large in the near horizon
region, possibly leading to either $b_{hor}\gg1$ or $b_{hor}\ll1$. In either
of these cases $\mathcal{R}_{\max}\rightarrow1$, indicating an infrared
radionic reflectivity.

\section{Summary}

A Kaluza-Klein model with two spherically compactified extra dimensions,
studied previously by Davidson and Guendelmann\cite{DG1,DG2} and by Carroll,
Geddes, Hoffman, and Wald\cite{CGHW}, is examined here with attention focusing
on the development of a radion cloud around an evaporating neutral,
nonrotating MBH. The cloud owes its existence not to the gravitational field
alone (a Schwarzschild solution is accompanied by a trivial radion solution
$\varphi=0$), but arises in response to the environmental effect of the
Hawking radiation. The radiation is modeled as a fluid with an effective
energy density $\rho(r,t)$ contributing to the radion equation of motion. An
effective pressure $p$ is assumed to be related to $\rho$ through an equation
of state $p/\rho=w=$ const. with $0\leq w\leq1$, resembling the perfect fluid
Hawking \textquotedblleft atmosphere\textquotedblright\ models of refs.
\cite{Zurek-Page} and \cite{'t Hooft}. For the particular case $w=1/3$, as is
expected for a fluid of noninteracting masseless particles in thermal
equilibrium, there is no environmental effect on $\varphi$. However, it is not
assumed here that the fluid is in equilibrium, and the particle modes
contributing to the energy density $\rho=\rho_{\text{tot}}-\rho_{0}$ (where
$\rho_{0}$ is due to massless particle modes) are those associated with
particles of nonzero rest mass. The matter contribution to the radion
effective potential is $U_{matter}=-\mathcal{T}_{cl}\ln b=$ $-\lambda\zeta\ln
b$, which can be positive, negative, or zero, depending on the sign of
$\zeta=(1-3w)\rho/\lambda$. A classical description has been used to estimate
the $\sigma=\left\langle \partial\mathcal{L}_{m}/\partial\varphi\right\rangle
$ term in the radion equation of motion, but near the horizon quantum field
effects such as vacuum polarization\cite{DFU},\cite{CF},\cite{FZ} are expected
to be important and may contribute to a shift in the radion vacuum.

The radion $\varphi$ approaches a normal vacuum value $\varphi\rightarrow0$
($b\rightarrow1$) asymptotically, where the Hawking radiation energy density
vanishes, but near the MBH the radion is shifted to a value $\varphi\neq0$
($b\neq1$) for any $\zeta\neq0$. Gradients of $\varphi$ and a nonzero radion
potential $V(\varphi)$ then give rise to a radion cloud with nonvanishing
energy density around the MBH. This radion cloud has an estimated size
$R(t)\sim\left(  \frac{M_{0}}{M(t)}\right)  m_{\varphi}^{-1}$ and an energy
dependent reflection coefficient $\mathcal{R}(\omega)$ as studied
in\cite{NDL-JM}. This reflection coefficient has a maximum value
$\mathcal{R}_{\max}$ (given by (\ref{a1}) for the case of electromagnetic
radiation in the flat space limit), which depends upon the parameter $\zeta$
near the horizon. $\mathcal{R}(\omega)/\mathcal{R}_{\max}$ begins to become
significant for particle energies $\omega\lesssim R^{-1}(t)$. Since the
asymptotic compactification radius for the extra dimensions is $\sim
m_{\varphi}^{-1}$, the size of the cloud compared to that of the extra
dimensions in asymptotic space is $R(t)/m_{\varphi}^{-1}\sim M_{0}/M(t)$ which
is initially small, but becomes of order unity at the end stages of the
evaporation. An infrared portion of the Hawking spectrum detected by an
external observer will be suppressed if $\mathcal{R}_{\max}\rightarrow1$, and
some low energy particles incident upon the MBH from the outside will be
reflected back. The amount of reflectivity depends upon the temperature $T$ of
the MBH (and therefore the parameter $\zeta$ near the horizon) and particle
energy $\omega$. For high energy particles ($\omega\gg R^{-1}$) the radion
cloud is transparent. The Hawking radiation contributes heavily to the
effective potential $U(\varphi)$ for large $|\zeta|$, in which case
$\mathcal{R}_{\max}$ may approach unity. For $\zeta<0$ one may have a vacuum
with $b_{hor}\ll1$, while for positive values $\zeta\gtrsim.5$, the effective
potential $U(\varphi)$ is completely destabilized, i.e., a local minimum
disappears. In this case the slope of $U$ is negative, and the radion rolls
outward in time with $b(t)$ increasing. In either case, when $\zeta\neq0$, a
radion cloud must develop, since $\varphi=0$ and $b=1$ is not a solution of
the radion EoM. For $\zeta\neq0$ the radion infrared albedo effect increases
as the MBH evaporates. Since transparency begins to set in at particle
energies $\omega\gtrsim R^{-1}\sim\left(  \frac{M}{M_{0}}\right)  m_{\varphi
}\gtrsim m_{\varphi}$, the energy range of observable albedo effects
($\omega\lesssim m_{\varphi}$) will be very sensitive to the radion mass
$m_{\varphi}$ and therefore the size of the extra dimensions.

\medskip

\textbf{Acknowledgement:} \ I thank Eduardo Guendelman and an anonymous
referee for useful comments.

\end{document}